\documentclass[11pt,twoside]{article}
\usepackage{asp2006}
\usepackage{graphics}
\usepackage{graphicx}
\begin{document}
\title{Surface Flows From Magnetograms}
\author{B.~T. Welsch\altaffilmark{1} and G.~H. Fisher\altaffilmark{1}}

\altaffiltext{1}{Space Sciences Laboratory, University of California, 
7 Gauss Way, Berkeley, CA 94720-7450}

\begin{abstract}
Estimates of velocities from time series of photospheric and/or
chromospheric vector magnetograms can be used to determine fluxes of
magnetic energy (the Poynting flux) and helicity across the
magnetogram layer, and to provide time-dependent boundary conditions
for data-driven simulations of the solar atmosphere above this layer.
Velocity components perpendicular to the magnetic field are necessary
both to compute these transport rates and to derive model boundary
conditions.  Here, we discuss some possible approaches to estimating
perpendicular flows from magnetograms.  Since Doppler shifts contain
contributions from flows parallel to the magnetic field, perpendicular
velocities are not generally recoverable from Doppler shifts alone.
The induction equation's vertical component relates evolution in $B_z$
to the perpendicular flow field, but has a finite null space, meaning
some ``null'' flows, e.g., motions along contours of normal field, do
not affect $B_z$.  Consequently, additional information is required to
accurately specify the perpendicular flow field.  Tracking methods,
which analyze $\partial_t B_z$ in a neighborhood, have a long
heritage, but other approaches have recently been developed.  In a
recent paper, several such techniques were tested using synthetic
magnetograms from MHD simulations.  Here, we use the same test data to
characterize: 1) the ability of the induction equation's normal
component, by itself, to estimate flows; and 2) a tracking method's
ability to recover flow components that are perpendicular to
$\mathbf{B}$ and parallel to contours of $B_z$.  This work has been
supported by NASA Heliophysics Theory grant NNG05G144G.
\end{abstract}

\section{Why study surface flows from magnetograms?}
\label{sec:intro}

The large length scales and relatively high conductivity of the plasma
in the solar corona imply that, to a good approximation, magnetic flux
is frozen to the plasma there.  Consequently, the coronal magnetic
field is ``line-tied'' to the plasma in lower atmospheric layers where
hydrodynamic forces can be stronger than Lorentz forces --- the
photosphere and lower chromosphere --- and coronal evolution is
strongly coupled to evolution in these layers.  Accordingly,
observations of magnetic field evolution below the Sun's corona ---
typically, sequences of photospheric or chromospheric magnetograms ---
provide crucial tools understand coronal evolution.

Usually, vector magnetograms are more useful than line-of-sight (LOS)
magnetograms for studying coronal evolution, because information
derived from LOS measurements alone will not, in general, be
consistent with the actual magnetic field, which has field components
both parallel and transverse to the LOS.  Although time series of
vector magnetograms have historically been rare, SOLIS
\cite{BW_Henney2002}, the Solar Optical Telescope (SOT, Tarbell 2006)
\nocite{BW_Tarbell2006} on Hinode, and the Solar Dynamics Observatory's
Helioseismic and Magnetic Imager (HMI) \cite{BW_Scherrer2005}, should
dramatically improve photospheric vector magnetogram spatial and
temporal coverage in the near future.

Several techniques have been developed to derive flows from time
series of magnetograms \cite{BW_Chae2001,BW_Kusano2002,BW_Welsch2004,
BW_Longcope2004,BW_Georgoulis2006,BW_Schuck2006}.
Estimated flows at the base of the corona can be used to derive the
fluxes of magnetic helicity, energy, and free energy into the corona,
\cite{BW_Demoulin2003,BW_Pariat2005,BW_Welsch2006}.  Further, flow estimates
can be used to provide time-dependent boundary conditions for
data-driven simulations of coronal magnetic field evolution.

This paper briefly reviews progress on estimating surface flows from
magnetogram sequences, and demonstrates some aspects of the problem 
with test data.

\section{Progress in estimating surface flows from magnetograms}

How have velocities been derived from vector magnetograms?  Chae (2001)
\nocite{BW_Chae2001} applied local correlation tracking (LCT)
\cite{BW_November1988} to LOS photospheric magnetograms to determine the
proper motions of magnetic features on the magnetogram surface, and
assumed the inferred flows $\mathbf{u}^{(\rm LCT)}$ were estimates of the
horizontal plasma velocities $\mathbf{v}_h$.  (Here the $h$ subscript
denotes a vector's components tangential to the magnetogram surface.
We avoid the subscript $t$, which has been used to refer to vector
components both transverse to the LOS and tangential to the solar
surface.)  Assuming that the observed evolution of the photospheric
magnetic field is governed by flows according to the ideal induction
equation,
\begin{equation}
\frac{\partial \mathbf{B}}{\partial t} 
= - c \, (\nabla \times \mathbf{E}) 
= \nabla \times (\mathbf{v} \times \mathbf{B}) 
~, \label{eqn:induct} \end{equation} 
Kusano {\it et al.}~(2002) \nocite{BW_Kusano2002} proposed using the
component of equation (\ref{eqn:induct}) normal to the magnetogram,
\begin{equation}
\frac{\partial B_z}{\partial t}
= \hat{\mathbf z} \cdot \nabla \times (\mathbf{v} \times \mathbf{B}) 
= - \nabla \cdot (\mathbf{v}_h B_z - v_z \mathbf{B}_h)
~, \label{eqn:normal} \end{equation} 
to derive three-component velocity fields --- $v_x, v_y, v_z$.  Here,
we have defined the magnetogram surface to be the horizontal plane
containing $\hat{\bf x}$ and $\hat{\bf y}$, with a vertical normal
$\hat{\bf z}$.  The three components of $\mathbf{v}$ cannot completely
determined from (\ref{eqn:normal}) alone, so more data or assumptions
are required to close the system for $\mathbf{v}$.

What about the horizontal components of equation (\ref{eqn:induct})?
As Kusano {\it et al.}~(2002) \nocite{BW_Kusano2002} noted, only the normal
component of equation (\ref{eqn:induct}) is completely specified by
vector magnetic field measurements from a single atmospheric layer;
the other components of equation (\ref{eqn:induct}) contain vertical
derivatives of horizontal magnetic field components, and therefore
require measurements of the vector magnetic field at a different
height in the atmosphere (e.g., the chromosphere) which are only
rarely available \cite{BW_Leka2003c,BW_Metcalf2005}.  

Kusano {\it et al.}~(2002) \nocite{BW_Kusano2002} assumed LCT velocities, $\mathbf{u}$, to be
equivalent to $\mathbf{v}_h$ to close the system.  D\'emoulin and
Berger (2003) \nocite{BW_Demoulin2003} argued that tracked motions of
magnetic flux on the solar photosphere, $\mathbf{u}$, result from the
combined effects of horizontal plasma velocities transporting vertical
magnetic fields and vertical plasma velocities transporting horizontal
magnetic fields, via
\begin{equation} \mathbf{u} B_z = \mathbf{v}_h B_z - v_z \mathbf{B}_h
~. \label{eqn:dandb} \end{equation}
The distinction between apparent motions of flux and plasma velocities
led Welsch (2006) to term $\mathbf{u}$ the ``flux transport velocity.''
D\'emoulin and Berger (2003) \nocite{BW_Demoulin2003} suggested that LCT
could be used to infer $\mathbf{u}$, but not $\mathbf{v}_h$ directly.  We
note that, in addition to ideal flux transport, diffusive effects can
also lead to apparent flux transport velocities \cite{BW_Welsch2006},
although we ignore these effects in the present work. 

Since the seminal work of Kusano {\it et al.}~(2002),
\nocite{BW_Kusano2002} still more techniques have been developed that
determine velocities from vector magnetograms.  Welsch {\it et
al.}~(2004) \nocite{BW_Welsch2004} used equation (\ref{eqn:dandb}) to
combine Fourier LCT (FLCT) results with equation (\ref{eqn:normal}) to
determine a photospheric flow field in a method they termed inductive
local correlation tracking, or ILCT.  Longcope (2004)
\nocite{BW_Longcope2004} developed the minimum energy fit (MEF), which
finds the photospheric velocity field that is strictly consistent with
equation (\ref{eqn:normal}) and that minimizes a penalty function,
e.g., the integrated square of the three-component photospheric
velocity.  Georgoulis and LaBonte (2006) \nocite{BW_Georgoulis2006}
extended the minimum structure method of Georgoulis, LaBonte, and
Metcalf (2004) \nocite{BW_Georgoulis2004} to the problem of velocity
determination, in a method they termed minimum-structure
reconstruction (MSR).  Schuck (2005) \nocite{BW_Schuck2005} showed that,
formally, LCT is not consistent with the induction equation's normal
component, which can be expressed as a continuity equation; instead,
LCT is consistent with the advection equation.  Building upon the
``differential LCT'' (DLCT) method developed by Lucas and Kanade
(1981), \nocite{BW_Lucas1981} Schuck (2006) developed the differential
affine velocity estimator (DAVE), which employs least-squares fitting
to solve the continuity-form of equation (\ref{eqn:normal}) for
$\mathbf{u}$ and its spatial derivatives. \nocite{BW_Schuck2006}  All of
the methods listed above can be applied to chromospheric as well as
photospheric magnetograms.

Flow estimation techniques that explicitly employ equation
(\ref{eqn:normal}) have been termed ``inductive''
\cite{BW_Georgoulis2006}; examples include IM, ILCT, MEF, and MSR.
Tracking methods, e.g., FLCT, DLCT, and DAVE, use the evolution of
structure in $B_z$ to quantify proper motions, and may also be
referred to as ``optical flow'' techniques \cite{BW_Schuck2006}.
Tracking techniques usually apply a windowing function, centered on
each pixel tracked, to derive the optical flow in the neighborhood of
that pixel.  Since this window imposes a scale length, tracking
methods have been criticized for their selective sensitivity to flows
on the imposed window scale \cite{BW_Georgoulis2006}.  The tracking
techniques mentioned thus far might be termed ``Eulerian methods,'' as
they estimate velocities over a set of pixels from an image pair. In
contrast, ``feature tracking'' techniques \cite{BW_DeForest2007}, in
which discrete ``features'' are identified from structure in $B_z$
maps and followed in time, might be termed ``Lagrangian'' techniques.
We will not discuss feature tracking techniques further in this work.

Generally, velocity estimation techniques are susceptible to errors in
magnetograms, because changes in the inferred magnetic field,
$\partial_t B_z$, are assumed to arise from flows via equation
(\ref{eqn:normal}).  Simply put, when noise or systematic errors
introduce spurious fluctuations $\delta B_z$ in $B_z$, spurious
velocities are derived.  In particular, in regions where $|B_z|$ is
small, $\delta B_z$ due to noise can be on the order of changes
$\Delta B_z$ due to actual field evolution.  For this reason, tracking
methods are expected to estimate flows more accurately in regions
where $|B_z|$ is large (barring magnetograph saturation effects) than
where $|B_z|$ is small.

The recent proliferation of velocity estimation methods led Welsch
{\it et al.}~(2007) \nocite{BW_Welsch2007} to test seven routines'
ability to reproduce known flows.  They extracted several pairs of
``synthetic magnetograms'' from ANMHD simulations of a rising flux rope
in the upper solar convection zone, and used several methods
methods --- two LCT codes, DAVE, IM, ILCT, MEF, and MSR --- to
estimate flows responsible for magnetic evolution between each pair.
They then compared properties of the estimated flows with those of the
``true'' flows from the MHD code.  (These flow reconstructions were
not blind: those deriving flow estimates had access to the true flow.)
They found that MEF, DAVE, FLCT, IM, and ILCT performed similarly by
many measures, but that MEF estimated the fluxes of magnetic energy
and helicity quite well.  The other methods tested estimated the
fluxes of magnetic energy and helicity through the magnetogram layer
poorly.

While the MHD simulations of field evolution in the solar interior
used by Welsch {\it et al.}~(2007) \nocite{BW_Welsch2007} did not
accurately model photospheric field evolution, they did provide a
valuable tool for testing velocity estimates.  In the near future,
blind tests, with simulated data that more accurately model
photospheric field evolution, are planned, as are tests of the
sensitivity of flow estimation methods to noise.  (Application of
these methods to synthetic chromospheric vector magnetograms will also
be investigated.)

\section{Theoretical considerations}

As noted in the discussion of equation (\ref{eqn:normal}), $\partial_t
B_z$ alone cannot fully specify the three components of $\mathbf{v}$.
We note that both evolution in $\mathbf{B}$ (equation [\ref{eqn:induct}]) and 
the ideal, upward fluxes of magnetic energy $S_z$ \cite{BW_Demoulin2003}, 
\begin{equation} 
S_z = \frac{-1}{4\pi} \int d\!A' \, 
\hat{\mathbf{z}} \cdot [(\mathbf{v} \times \mathbf{B}) \times \mathbf{B}] 
= \frac{-1}{4\pi} \int d\!A' \, 
(\mathbf{u} B_z) \cdot \mathbf{B}_h   
~, \label{eqn:dSdt}   \end{equation}
and magnetic helicity (D\'emoulin and Berger 2003; see also Pariat
{\it et al.}~  2005) \nocite{BW_Demoulin2003} \nocite{BW_Pariat2005}
\begin{equation} 
dH_A/dt = 2 \int d\!A' \, 
\hat{\mathbf{z}} \cdot [\mathbf{A}_P \times (\mathbf{v} \times \mathbf{B}) ]
= - 2 \int d\!A' \, 
(\mathbf{A}_P \cdot \mathbf{u}) B_z
 \label{eqn:dhdt_bf} \end{equation}
are all linear in $(\mathbf{v} \times \mathbf{B})$.  (In equation
[\ref{eqn:dhdt_bf}], $\nabla_h = [\partial_x, \partial_y, 0]^T$,
$\hat{\mathbf z} \cdot \nabla_h \times \mathbf{A}_P = B_z$ and
$\nabla_h \cdot \mathbf{A}_P = \hat{\mathbf z} \cdot \mathbf{A}_P =
0$.)  This dependence means that flows along the magnetic field,
$v_\parallel$, do not lead to any evolution in $\mathbf{B}$, and do
not transport either magnetic energy or magnetic helicity across the
magnetogram surface.  Hence, the components of $\mathbf{v}$
perpendicular to $\mathbf{B}$,
\begin{equation} \mathbf{v}_\perp \cdot \mathbf{B} = 0 
~, \label{eqn:vB0}  \end{equation}
are of practical interest for most applications of velocity
estimation. This equation, combined with equation
(\ref{eqn:normal}), now provides two equations for the three unknown
components of $\mathbf{v}$.  

In this section, we briefly consider possible remedies to the
remaining underdetermination, as well as its implications.

\subsection{Can Doppler data determine surface magnetic flows directly?}

\begin{figure}[!ht]
 \includegraphics[width=5.0in]{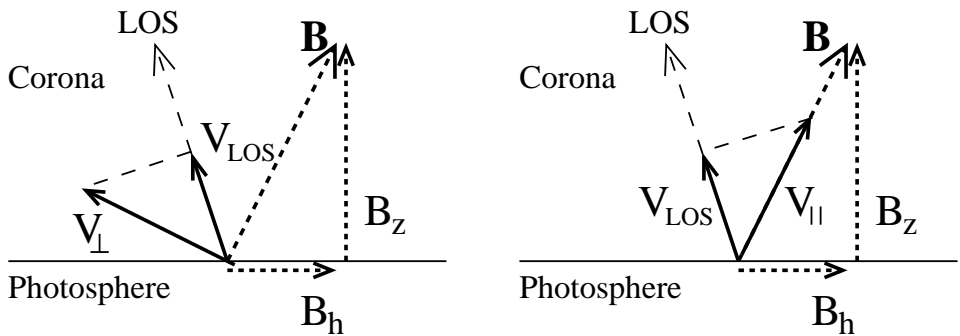}
  \caption{When $B_{\rm LOS} \ne 0$, a flow $\mathbf{v}_\perp$
perpendicular to $\mathbf{B}$ can produce the same LOS velocity as a
flow $\mathbf{v}_\parallel$ parallel to $\mathbf{B}$.  This means LOS
flows inferred from Doppler shifts are insensitive to the relative
orientation of $\mathbf{v}$ and $\mathbf{B}.$ Flows parallel to
$\mathbf{B}$ do not lead to evolution in $\mathbf{B}$, and do not
transport magnetic energy or helicity.  The evolution of $B_z$ at the
photosphere, however, can be used to estimate $\mathbf{v}_\perp.$ The
presence of a field component perpendicular to the page (neglected in
this figure) would not affect this analysis.  }
\label{fig:dopdegen}
\end{figure}

Doppler measurements of the velocity of the magnetized plasma, best
determined from Stokes $V,Q,$ and/or $U$ profiles (not $I$; see Chae
{\it et al.}~[2004]), \nocite{BW_Chae2004} cannot fully determine whether the
inferred plasma flows lie either parallel to or perpendicular to the
magnetic field if $B_{\rm LOS} \ne 0$.  This degeneracy is illustrated
schematically in Figure \ref{fig:dopdegen}.

Doppler shifts do unambiguously determine the perpendicular flow when
$B_{\rm LOS} = 0$.  Near disk center, this is approximately satisfied
along polarity inversion lines, where $B_z$ changes sign, as noted by
\cite{BW_Chae2004} and \cite{BW_Lites2005}.  Doppler observations of
have shown that some polarity inversion lines can also be velocity
inversion lines (Deng {\it et al.}~2006), \nocite{BW_Deng2006} which can
be interpreted as the manifestation of siphon flows moving along field
lines that arch over the polarity inversion line.

Doppler measurements can, however, be combined with an estimate of the
perpendicular velocity $\mathbf{v}_\perp$ from one of the techniques
described above to recover the flow parallel to the magnetic field:
subtracting the projection of $\mathbf{v}_\perp$ onto the LOS from
$v_{\rm LOS}$ gives the LOS component of $v_\parallel$, which can be
divided by the cosine of the angle between $\mathbf{B}$ and the LOS to
give $v_\parallel$.  Georgoulis and LaBonte (2006)
\nocite{BW_Georgoulis2006} have employed this approach in an
observational study of active region flows, and Ravindra and Longcope
(2007) \nocite{BW_Ravindra2007} have investigated the use of Doppler
shifts in a theoretical study using the data from Welsch {\it et
al.}~(2007). \nocite{BW_Welsch2007}

In the following discussion, any reference to $\mathbf{v}$ should
be assumed to refer to $\mathbf{v}_\perp$.

\subsection{Inductive flows}
\label{subsec:induct}

One can employ a Helmholtz decomposition of equation (\ref{eqn:dandb}),
\begin{equation} 
\mathbf{u} B_z = \mathbf{v}_h B_z - v_z \mathbf{B}_h 
= -\nabla_h \chi - \nabla_h \times \psi \hat{\bf z} 
~, \label{eqn:decomp} \end{equation}
to express $\mathbf{u} B_z$ in terms of ``inductive'' and
``electrostatic'' potentials \cite{BW_Longcope2004} $\chi$ and $\psi$,
respectively.  (We note that, assuming the ideal Ohm's Law, these potentials
can be expressed as sources of a horizontal electric field,
\begin{equation} 
\mathbf{E}_h = c^{-1} (\hat{\mathbf z} \times \mathbf{u} B_z )
= c^{-1} ( \nabla_h \times \chi \hat{\bf z}) - c^{-1} \nabla_h \psi  
~.) \label{eqn:efield} \end{equation} 

Inserting equation (\ref{eqn:decomp}) into equation (\ref{eqn:normal})
yields a Poisson equation for $\chi$,
\begin{equation} 
\partial_t B_z = \nabla_h^2 \chi ~, \label{eqn:chi_poisson}
\end{equation} 
meaning the evolution of $B_z$ between a pair of sequential
magnetograms, $\Delta B_z/\Delta t$, specifies $\chi$.  This does not, 
however, constrain $\psi$ in any way.

Equation (\ref{eqn:chi_poisson}) can be solved to determine the purely
inductive flux transport rate, i.e., one that assumes $\psi$ = 0,
\begin{equation} 
\mathbf{u}_I B_z = - \nabla_h \chi 
~, \label{eqn:uibz} \end{equation} 
where the subscript $I$ denotes that the flux transport velocity is
inductive.  In principle, equation (\ref{eqn:uibz}) can be substituted
into equation (\ref{eqn:dandb}) and combined with equation
(\ref{eqn:vB0}) to determine to determine $\mathbf{v}$.  In practice,
though, solving this system of equations for $\mathbf{v}$ requires
division by $B_z$, which weights regions of weak $|B_z|$ more strongly
than regions of large $|B_z|$, and, as discussed in \S
\ref{sec:intro}, flows are expected to be poorly estimated in such
regions.  For some applications, however, an estimate of $\mathbf{u}
B_z$ is sufficient; for instance, equations (\ref{eqn:dSdt}) and
(\ref{eqn:dhdt_bf}) imply that $\mathbf{u} B_z$ can be used to
estimate the fluxes of magnetic energy and helicity into the corona
without determining $\mathbf{v}$.

\begin{figure}[!ht]
\includegraphics[width=5.0in]{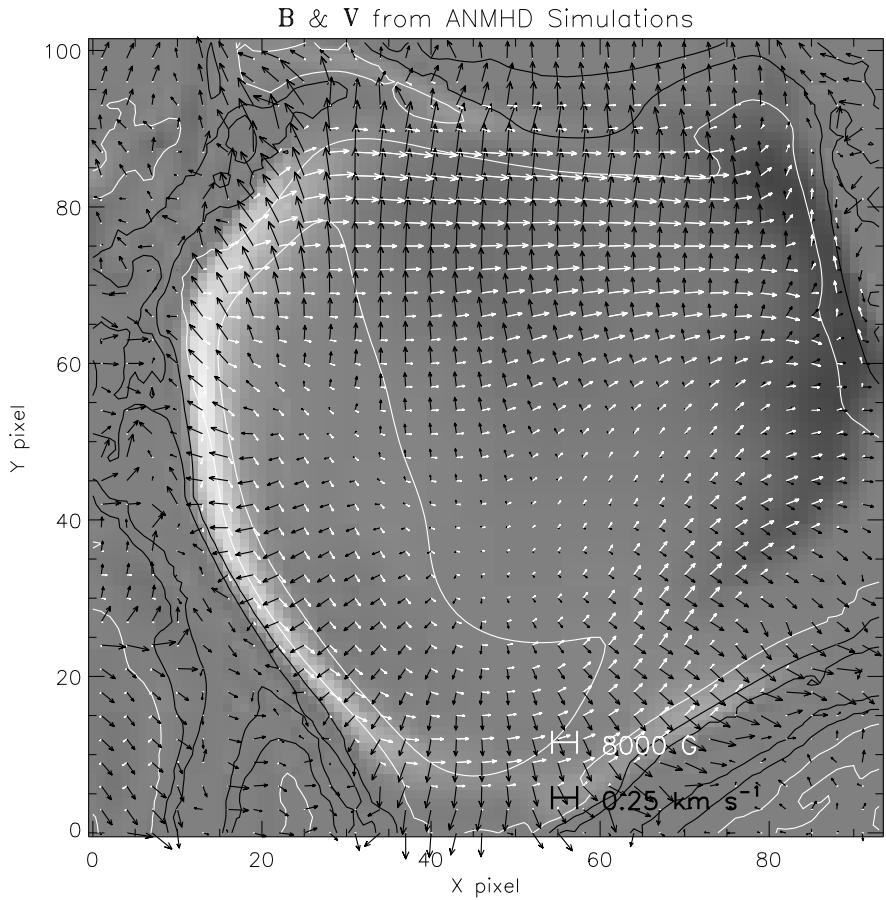}
  \caption{A snapshot of the magnetic and velocity fields in a cross
section of an ANMHD simulation of a flux rope rising through the
convection zone, extracted from near the top of the domain.  The
grayscale background shows $B_z$, white vectors show
$\mathbf{B}_h$, black (white) contours show smoothed downflows
(upflows) of $v_z$, and black arrows show $\mathbf{v}_h$.  Only
components of $\mathbf{v}$ perpendicular to $\mathbf{B}$ are shown. }
\label{fig:data}
\end{figure}

The ANMHD data used by Welsch {\it et al.}~(2007) \nocite{BW_Welsch2007}
can be used to demonstrate the ability to estimate flow properties
from the induction equation alone.  In Figure \ref{fig:data}, we show
a snapshot of the magnetic (grayscale background for $B_z$ and white
vectors for $\mathbf{B}_h$) and velocity (black/white contours for
$v_z$, and black arrows arrows for $\mathbf{v}_h$) fields in a cross
section of the simulation domain near the top of the convection zone.
We note that the strong fields present ($\sim 8000$ G) are appropriate
to the high-$\beta$ solar interior, but are substantially stronger
than observed photospheric field strengths.  These data are described
in much greater detail in Welsch {\it et al.}~  2007 \cite{BW_Welsch2007}.

\begin{figure}[!ht]
\includegraphics[width=5.0in]{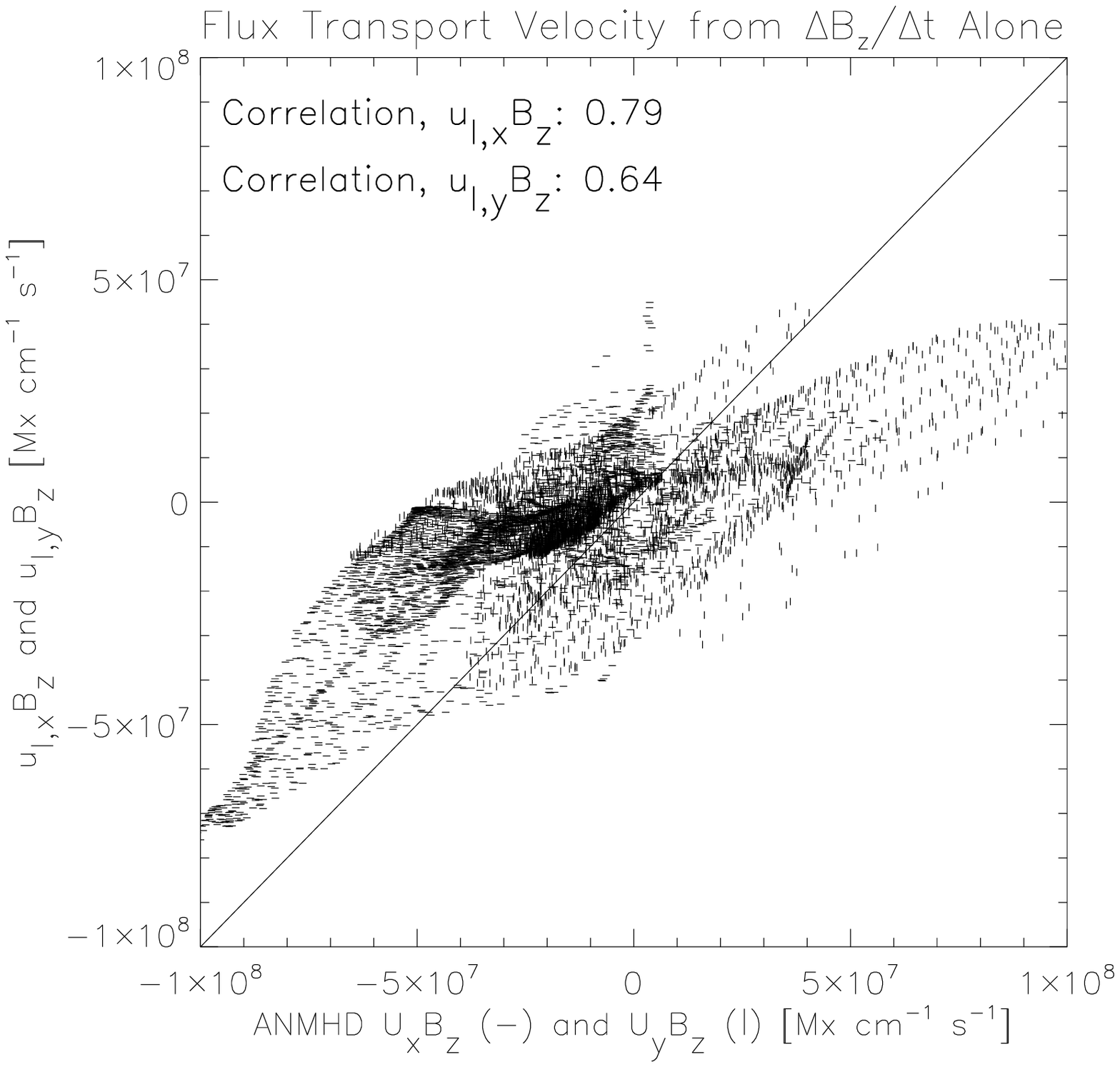}
  \caption{A scatter plot comparing ANMHD's flux transport rate,
    $\mathbf{U}_IB_z$ , with that derived from equation
    (\ref{eqn:chi_poisson}), $\mathbf{u}_IB_z$, along with the linear
    correlation coefficients for the $x$ and $y$ components.  The
    solid line is not a fit; it is the desired slope, and is shown to
    emphasize departures from that slope.}
\label{fig:uibz}
\end{figure}

Using the $\Delta B_z$ between two synthetic magnetograms 125 seconds
before and after the magnetogram shown in Figure \ref{fig:data}, we
solved equation (\ref{eqn:chi_poisson}) with a Fourier technique to
compute $\mathbf{u}_I B_z$.  Figure \ref{fig:uibz} shows a scatter
plot comparing ANMHD's flux transport rate, $\mathbf{U}_I B_z$, with
$\mathbf{u}_I B_z$, along with the linear correlation coefficients for
the $x$ and $y$ components.
These correlation coefficients compare quite favorably with results of
the tests by Welsch {\it et al.}~(2007). \nocite{BW_Welsch2007} We also
computed a least-absolute-difference linear fit of $|\mathbf{u}_IB_z|$
as a function of $|\mathbf{U}B_z|$, and found a slope of 0.4, when 1.0
would be ideal; evidently, inductive flux displacements underestimate
ANMHD's true flux displacements.

In addition, we calculated the estimated fluxes of magnetic energy and
magnetic helicity, separately trying the methods of D\'emoulin and
Berger (2003) \nocite{BW_Demoulin2003} and Pariat {\it et al.}~(2005)
\nocite{BW_Pariat2005} for the helicity flux.  We found that
$\mathbf{u}_I B_z$ determined by our Fourier method recovered 30\% of
the energy flux, but only 10\% of the helicity flux.  If, however, the
estimated helicity flux was summed only over strong-field pixels ---
in which $|B_z|$ exceeded 5\% of max($|B_z|$) --- the D\'emoulin and
Berger (2003) \nocite{BW_Demoulin2003} approach recovered 20\% of the
helicity flux from all pixels, while the Pariat {\it et al.}~(2005)
\nocite{BW_Pariat2005} approach recovered 15\% of the helicity flux from
all pixels.  While these results compare favorably with most of the
velocity estimation methods tested by Welsch {\it et al.}~(2007)
\nocite{BW_Welsch2007}, they reveal that inductively determined flux
transport rates lack essential information about the fluxes of energy
and helicity --- presumably related to $\psi$ being unconstrained by
equation (\ref{eqn:chi_poisson}).

The inductive potential $\chi$ used in these test was determined with
a Fourier technique, which assumes periodicity, and solves for $\chi$
on all pixels, including those with weak vertical field.  We ran the
same tests with an alternative potential function, $\chi'$, which
satisfies homogeneous boundary conditions over a masked region that
includes only pixels with $|B_z| > 170$ G.  Derived by B. Ravindra and
D.W. Longcope, $\chi'$ was used for the inductive component of MEF
flows in the tests conducted by Welsch {\it et
al.}~(2007). \nocite{BW_Welsch2007} We found that, compared to flux
transport rates from $\chi$, flux transport rates derived from $\chi'$
were more poorly correlated with ANMHD's flux transport rates,
underestimated $|\mathbf{U}B_z|$ more severely, and recovered less of
the vertical fluxes of magnetic energy and helicity.  We note that MEF
overestimated $|\mathbf{U}B_z|$ in the tests performed by Welsch {\it
et al.}~(2007) \nocite{BW_Welsch2007}.  This implies that the superior
performance of the MEF approach at estimating fluxes of magnetic
energy and helicity in Welsch {\it et al.}~(2007) \nocite{BW_Welsch2007}
derives from MEF's specification of $\psi$.

\subsection{Sensitivity of flow estimation to $\nabla_h B_z$}
\label{subsec:null}

As D\'emoulin and Berger (2003) \nocite{BW_Demoulin2003} observed, the
underdetermination of $\mathbf{v}$ by equation (\ref{eqn:normal})
implies that there exists a class of flows that cause no evolution in
$B_z$, but that can inject large amounts magnetic energy and helicity.
Flows that do not alter $B_z$ satisfy
\begin{equation} 
\mathbf{v}_{h,0} B_z - v_{z,0} \mathbf{B}_h 
= \mathbf{u}_0 B_z = - \nabla_h \times \psi \hat{\bf z} 
~. \label{eqn:nullflow} \end{equation}
Because these flows lie in the null space of equation
(\ref{eqn:normal}), we call them ``null flows,'' and denote them with
a zero subscript.  While flows along contours of $B_z$ --- ``contour
flows'' --- are a well known subset of null flows, modelers have
employed other null flows; see, e.g., \cite{BW_Lynch2005}.  

Taking the curl of equation (\ref{eqn:decomp}) yields 
\begin{equation} 
\nabla_h \times \mathbf{u} B_z = \nabla_h^2 \psi 
~. \label{eqn:psi_poisson} \end{equation} 
A flux transport velocity, $\mathbf{u}^{(EST)}$, estimated from any
optical flow technique (e.g., DAVE or FLCT) can be used with
(\ref{eqn:psi_poisson}) to find $\psi$, and equation
(\ref{eqn:chi_poisson}) to find $\chi$.  This is the essence of ILCT
\cite{BW_Welsch2004}; but it goes a step further, and uses equations
(\ref{eqn:decomp}) and (\ref{eqn:vB0}) to determine $\mathbf{v}$.
Since, however, optical flow techniques ultimately depend on $\Delta
B_z/ \Delta t$, and flows associated with $\psi$ do not alter $B_z$,
this procedure is expected to be insensitive to such flows.

\begin{figure}[!ht]
\includegraphics[width=5.0in]{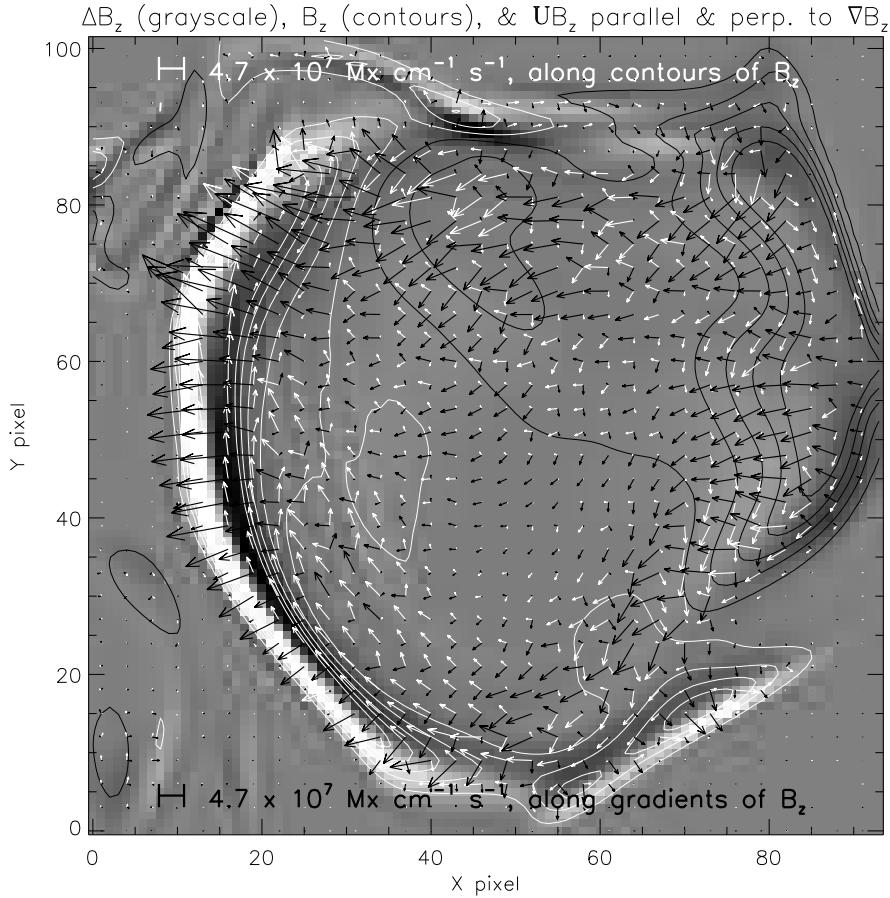}
  \caption{The change $\Delta B_z$ in the normal magnetic field is
shown in grayscale, along with selected contours of $B_z$ (white for
$B_z > 0$, black for $B_z < 0$).  ANMHD's instantaneous flux
displacement, $\mathbf{U} B_z$, decomposed into components along
contours (white vectors) and gradients (black vectors) of $B_z$, is
overplotted. Flow estimation techniques that depend upon $\partial_t
B_z$ are insensitive to flows that transport flux along contours of
$B_z$ (white vectors).  }
\label{fig:nullflow}
\end{figure}

We can use FLCT flows to illustrate the difference in sensitivity of
one velocity estimation technique to flows along contours of $B_z$
versus flows along gradients of $B_z$.  In Figure \ref{fig:nullflow},
we plot selected contours of ANMHD's average $B_z$ over a grayscale
image of the change $\Delta B_z$ over the same 250 second time
interval used to estimate $\mathbf{u}_I B_z$ in \S \ref{subsec:induct}
We also plot ANMHD's instantaneous flux displacement, $\mathbf{U}B_z$,
decomposed into components along contours (gradients) of of $B_z$ with
white (black) vectors.  Flow estimation techniques that depend upon
$\partial_t B_z$ are expected to be insensitive to flows that
transport flux along contours of $B_z$ (white vectors).

\begin{figure}[!ht]
\includegraphics[width=4.5in]{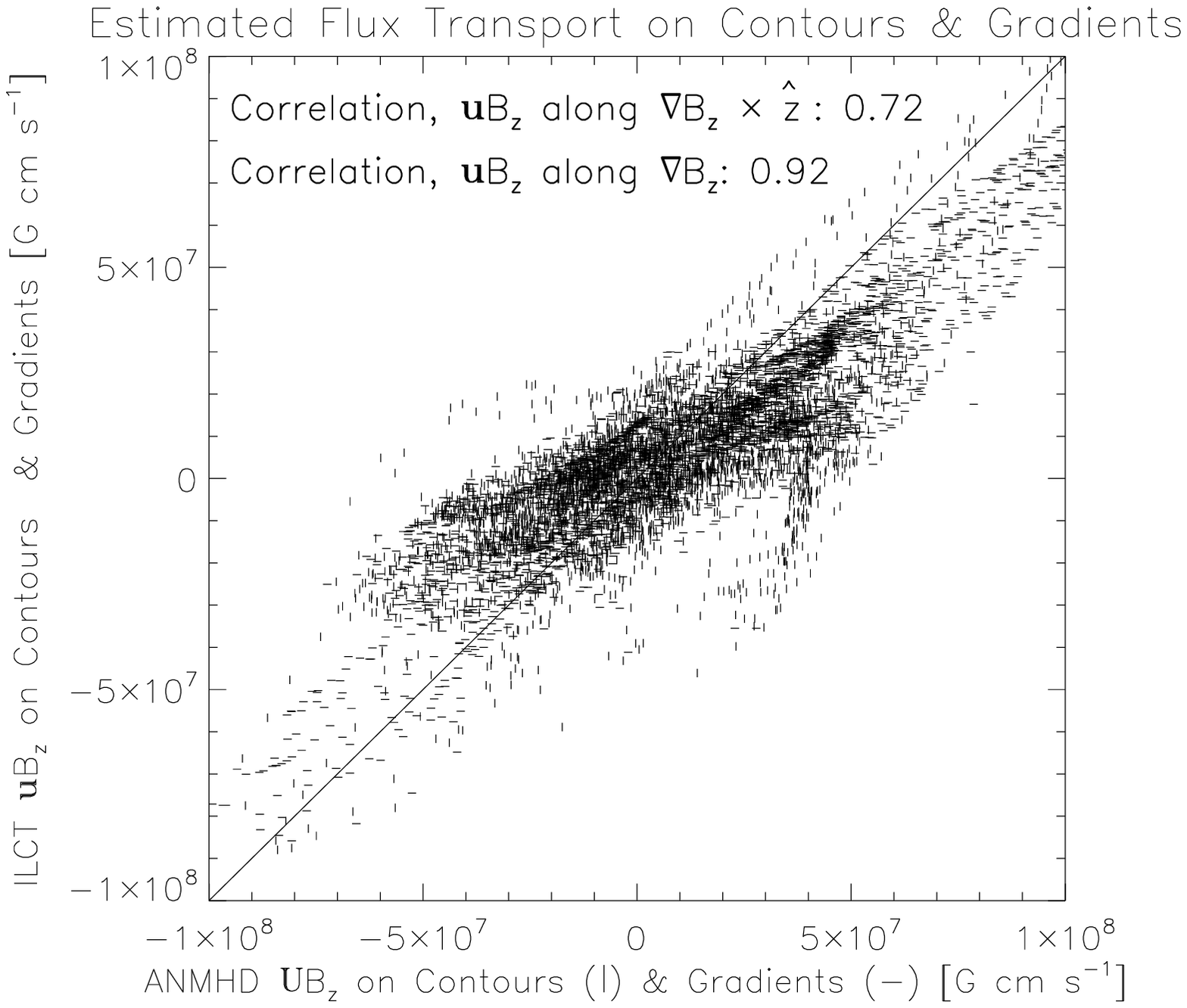}
  \caption{Scatter plots of ANMHD's $\mathbf{U} B_z$ 
    and estimated $\mathbf{u} B_z$ are shown.  The vectors
    have been decomposed into components along gradients in $B_z$
    (plotted with --)and contours of $B_z$ (plotted with $\vert$).  Flux
    transport along gradients of $B_z$ is recovered more accurately
    than transport along contours of $B_z$, as quantified by the
    linear correlation coefficients shown for each component.  The
    solid line is not a fit; it is the desired slope, and is shown to
    emphasize departures from that slope.}
\label{fig:contgradflow}
\end{figure}

Using $\mathbf{u}$ from FLCT in equation (\ref{eqn:psi_poisson}) to
find $\psi$, and equation (\ref{eqn:decomp}) to combine $\chi$ from
equation (\ref{eqn:chi_poisson}) with $\psi$, we attempted to
reconstruct ANMHD's flux displacement, $\mathbf{U} B_z$, from Figure
\ref{fig:nullflow}.  In Figure \ref{fig:contgradflow}, we show scatter
plots of flux displacements from ANMHD and our estimates of these
(denoted $\mathbf{U} B_z$ and $\mathbf{u} B_z$, respectively) along
$\nabla_h B_z$ (horizontal dashes) and contours of $B_z$ (vertical
line segments).  Flows along $\nabla_h B_z$ are recovered more
accurately than flows along contours of $B_z$, as quantified by the
linear correlation coefficients shown on the plot.  The fact that
contour flows are even partially recovered presumably results from
variations in the spatial structure of $B_z$ in the neighborhood of
each tracked pixel.

\section{Discussion}
\label{sec:discuss}

We have briefly reviewed central concepts regarding surface flow
estimation from magnetograms.  In addition, we have presented the
results of simple tests which demonstrate that the induction
equation's normal component, equation (\ref{eqn:normal}), can be used
to quantitatively estimate flux transport rates.  Fluxes of magnetic
energy and helicity derived from these estimates are, however, likely
to possess significant systematic errors.  We have also demonstrated
that flow estimation techniques that depend upon evolution in $B_z$
alone are insensitive to flux transport along contours of $B_z$,
compared to flux transport along $\nabla_h B_z$.  Equations
(\ref{eqn:dSdt}) and (\ref{eqn:dhdt_bf}), which depend on the dot
products of $\mathbf{u} B_z$ with $\mathbf{B}_h$ and with
$\mathbf{A}_P$ respectively, combined with the sensitivity of
$\mathbf{u} B_z$ to $\nabla_h B_z$, suggest that if $\mathbf{B}_h$
and/or $\mathbf{A}_P$ lie primarily along $\nabla_h B_z$, then the
fluxes of magnetic energy and/or helicity can, in principle, be
recovered accurately from $\Delta B_z/\Delta t$.  If, in contrast,
$\mathbf{B}_h$ and/or $\mathbf{A}_P$ lie primarily along contours of
$B_z$, then the fluxes of magnetic energy and/or helicity probably
cannot be recovered accurately from evolution in $B_z$ alone.

We note that, as discussed in \cite{BW_Welsch2007}, the ANMHD data used in the tests
presented here differ from actual magnetograms in significant ways, so
the properties of flows estimated from actual magnetograms will
probably differ substantially from the properties of flows estimated
from ANMHD data.

\acknowledgements Data from MEF velocity estimates were courtesy of
B. Ravindra and D.~W. Longcope.  We acknowledge support from grant
NNG05G144G-04/08, from NASA's Sun-Earth Connections Theory Program.

\end{document}